\PassOptionsToPackage{nopatch=footnote}{microtype}
\documentclass[sigconf]{acmart}
\AtBeginDocument{%
  }

\setcopyright{acmlicensed}
\copyrightyear{2026}
\acmYear{2026}
\acmDOI{XXXXXXX.XXXXXXX}
\acmConference[MobiSys '26]{The 24th Annual International Conference
  on Mobile Systems, Applications, and Services (Demo)}{June 21--25, 2026}{Cambridge, UK}
\acmISBN{978-1-4503-XXXX-X/2026/06}

\begin{document}

\title{Demo: Real-Time Cross-Layer Semantic Error Correction Using Language Models and Software-Defined Radio}

\author{Yuchen Pan}
\affiliation{%
  \institution{The Chinese University of Hong Kong}
  \city{Hong Kong SAR}
  \country{China}}
\email{ypan@link.cuhk.edu.hk}

\author{Yuyang Du}
\affiliation{%
  \institution{The Chinese University of Hong Kong}
  \city{Hong Kong SAR}
  \country{China}}
\email{yuydu@ie.cuhk.edu.hk}

\author{Yirun Wang}
\affiliation{
  \institution{The Chinese University of Hong Kong}
  \city{Hong Kong SAR}
  \country{China}}
\email{yrwang@ie.cuhk.edu.hk}

\author{Shiqi Xu}
\affiliation{
  \institution{The Chinese University of Hong Kong}
  \city{Hong Kong SAR}
  \country{China}}
\email{xs024@ie.cuhk.edu.hk}

\author{Lihao Zhang}
\affiliation{
  \institution{The Chinese University of Hong Kong}
  \city{Hong Kong SAR}
  \country{China}}
\email{lihaocuhk@link.cuhk.edu.hk}

\author{Soung Chang Liew}
\affiliation{%
  \institution{The Chinese University of Hong Kong}
  \city{Hong Kong SAR}
  \country{China}}
\email{soung@ie.cuhk.edu.hk}

\renewcommand{\shortauthors}{Pan, et al.}

\begin{abstract}
As Language Models (LMs) advance, Semantic Error Correction (SEC) has emerged as a promising approach for reliable network designs. Yet existing methods prioritize intent over accuracy, falling short of verbatim recovery. Our recent work, Cross-Layer SEC (CL-SEC), addressed this by fusing physical-layer Log-Likelihood Ratios (LLRs) with semantic context, but its real-time feasibility remained unvalidated. This paper demonstrates CL-SEC on a live  Software-Defined Radio (SDR) testbed, resolving implementation barriers with: 1) an SDR middleware enabling real-time LLR extraction from FPGA hardware, and 2) a generalized inference interface supporting modern encoder-decoder LMs. Real-world experiments confirm that the cross-layer fusion significantly outperforms either source alone.
\end{abstract}



\maketitle

\section{Introduction}

\begin{figure}[t]
  \centering
  \includegraphics[width=\linewidth]{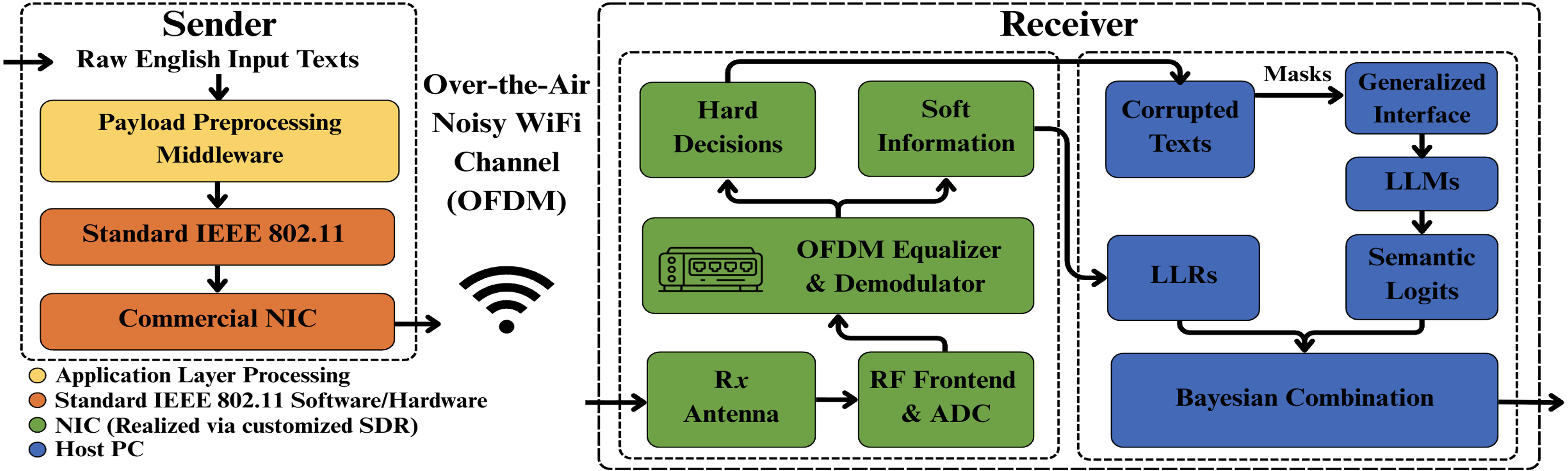}
  \vspace{-20pt}
  \caption{System architecture of our CL-SEC implementation.}
  \label{fig:overall}
  \vspace{-5pt}
\end{figure}

\begin{figure*}[t]
  \centering
  \begin{minipage}[t]{0.31\textwidth}
  \centering
  \raisebox{10pt}{\includegraphics[width=\linewidth]{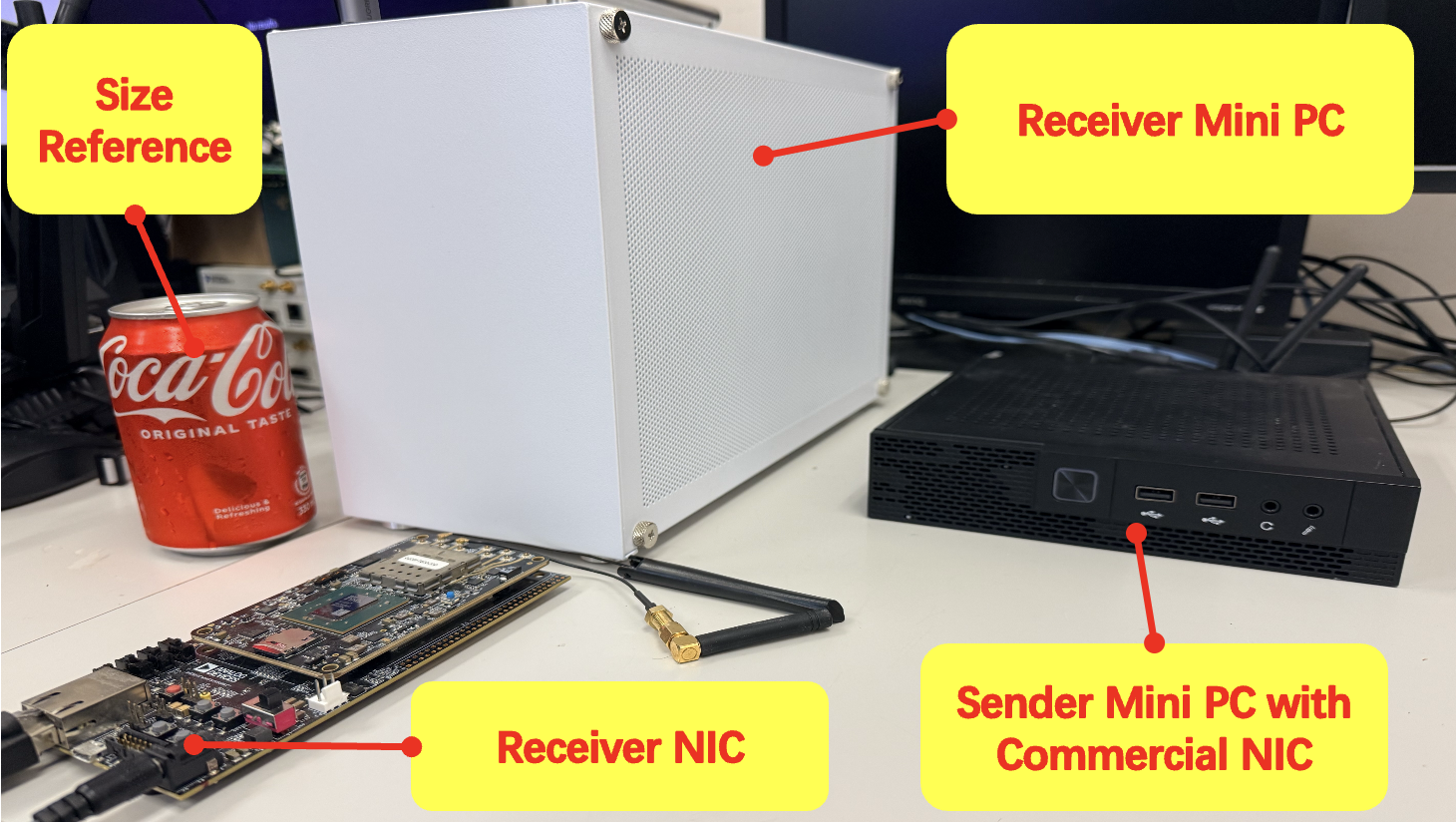}}
  \vspace{-28pt} 
  \caption{Real-world CL-SEC hardware testbed.}
  \label{fig:real}
\end{minipage}
  \hfill 
  \begin{minipage}[t]{0.31\textwidth}
    \centering
    \includegraphics[width=\linewidth]{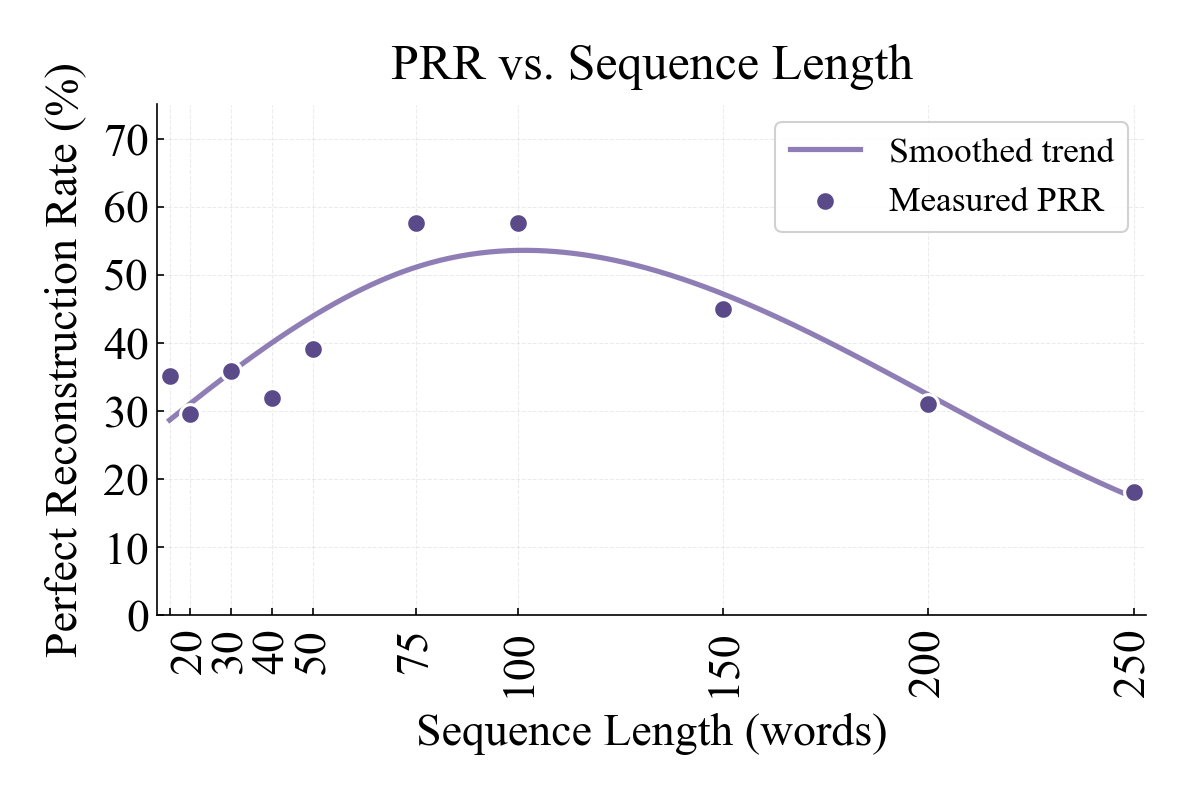}
    \vspace{-28pt}
    \caption{PRR vs. Sequence Length using LLR + T5Gemma}
    \label{fig:3}
  \end{minipage}
  \hfill
  \begin{minipage}[t]{0.31\textwidth}
    \centering
    \includegraphics[width=\linewidth]{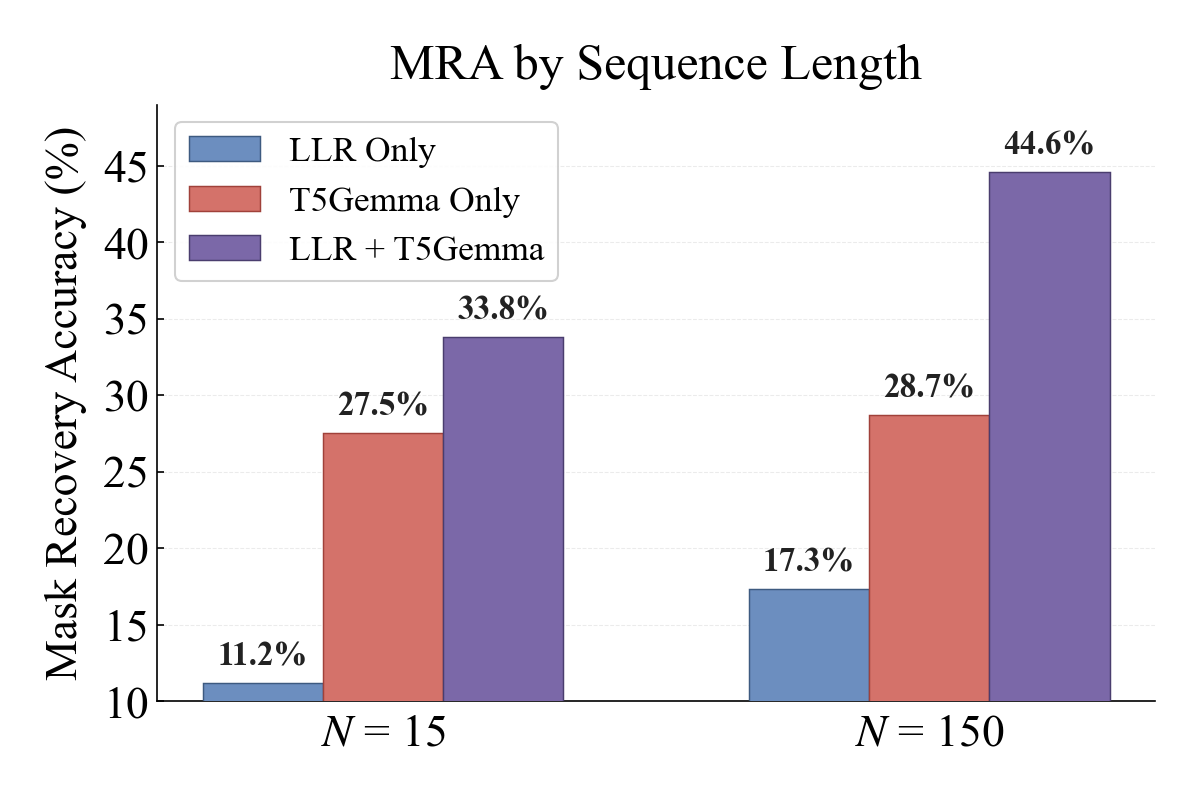}
    \vspace{-28pt}
    \caption{MRA for 15 and 150-word sequences across three configurations.}
    \label{fig:4}
  \end{minipage}
  \vspace{-6pt}
\end{figure*}

Semantic Error Correction (SEC) has recently emerged as a promising paradigm for enhancing communication reliability by leveraging the vast contextual knowledge and reasoning capabilities of Language Models (LMs). By prioritizing the recovery of a message's core intent rather than just its bitstream, SEC demonstrates remarkable resilience in low-SNR environments. However, existing SEC methods, such as \cite{hao2025short}, struggle with verbatim reconstruction. To bridge this gap, our recent work proposed cross-layer SEC (CL-SEC) \cite{Wang2026CL}, a theoretical framework that integrates physical-layer Log-Likelihood Ratios (LLRs) with LM-driven SEC at the application layer to enable precise, word-for-word message recovery.

Despite its potential, the initial CL-SEC framework in \cite{Wang2026CL} was validated solely through offline simulations, leaving its real-time feasibility on practical hardware an open problem. Additionally, the design in \cite{Wang2026CL} was restricted to encoder-only models, a limitation that diverges from the current trajectory of LM evolution, ultimately constraining the framework's long-term adaptability.

This paper provides a real-world demonstration of CL-SEC on Software-Defined Radio (SDR). We answer the problem of real-time feasibility through a plug-and-play and low-latency SDR middleware that extracts physical-layer data directly from hardware buffers and relays it to the application layer. This middleware effectively bypasses standard MAC-layer drop policies by operating transparently relative to higher network stacks. Furthermore, we introduce a generalized LM inference interface that unifies the processing of both encoder-only and encoder-decoder models. This design resolves previous architectural constraints, unlocking the potential of CL-SEC to work with the most advanced LMs today.\footnote{Design details of CL-SEC with the generalized LM interface, the SDR middleware, and our demo video are available at \textcolor{blue}{\url{https://github.com/Yirun719/CL-SEC}}, \textcolor{blue}{\url{https://github.com/Leo-Cheung-CUHK/SPECTER}}, and \textcolor{blue}{\url{https://youtu.be/X1imoPgvfeE}}, respectively.}

\section{Architecture Review and Implementation}
We begin with a brief review of the CL-SEC verbatim text recovery pipeline. Upon receiving a message, the system extracts hard-decision bits and LLRs from the physical-layer decoding process. These hard decisions are used to reconstruct a tentative text string, where any corrupted words identified through vocabulary checks are subsequently replaced with \texttt{<mask>} tokens. For each masked word, a LM generates semantic probabilities based on the contextual background within the application layer. Simultaneously, using the word's associated LLRs, CL-SEC generates physical-layer probabilities for potential corrections of the masked word. These two probabilities are then combined using a Bayesian product-form approach, and the candidate with the highest \textit{posterior }probability is selected as the final correction.

Fig. \ref{fig:overall} illustrates the framework of our real-world implementation. First, CL-SEC initializes the SDR device with the OpenWiFi firmware. A sender with commercial WiFi NIC connects to this SDR device and transmits text payloads via UDP. The SDR, serving as the receiver NIC, captures raw OFDM signals, extracts the underlying physical information, and streams the hard-decision bits and LLRs to the host PC. The framework incorporates two key designs:

\textbf{Customized SDR Middleware:} Standard MAC-layer processing discards soft physical-layer information before it reaches the application, making LLR extraction inaccessible on commodity hardware. To overcome this, we modified the FPGA-based OFDM receiver to compute a 20-bit LLR for each decoded bit and append these soft values onto the same DMA stream as the packet payload. A lightweight driver then parses this extended stream, extracting and exposing the LLR block to the application layer in real time, bypassing MAC-layer drop policies entirely and requiring no changes to the sender or network stack.

\textbf{Generalized LM Interface:} We further expanded our framework to integrate CL-SEC with modern encoder-decoder models. Unlike encoder-only architectures that natively perform single-word masked language modeling, encoder-decoder models generate outputs autoregressively, producing variable-length spans that do not naturally satisfy CL-SEC’s single-word constraint. To bridge this gap, we reformulate masked positions as span-infilling tasks: the encoder processes the masked sentence as context, while the decoder is primed to generate text for the target position. A structured sampling procedure then enforces word-boundary constraints, converting the autoregressive output into a ranked list of single-word candidates. This aligns the output with the processing logic of encoder-only models, ensuring full compatibility. See our GitHub link for implementation details.

\section{Experiments}
To validate the real-world performance of our framework, we build an SDR testbed in a noisy laboratory environment (Fig. \ref{fig:real}). The transmission hardware consists of a Sender Mini PC (\textit{Windows 11}) equipped with a commercial \textit{Qualcomm Atheros AR938x} Wi-Fi NIC. On the receiving end, the signal is captured by a \textit{ADI ADRV9361-Z7035} SDR device, which streams physical-layer data to a Mini PC (\textit{Ubuntu 20.04, RTX 2080 Ti}) for local real-time LM inference. 

We first evaluate how sequence length impacts the Perfect Reconstruction Rate (PRR)--the percentage of messages restored with 100\% word-for-word accuracy. The restoration relies on the Bayesian fusion of LLRs and semantic probabilities from T5Gemma-2b-2b-ul2, a modern encoder-decoder model. We use 10 datasets, each containing 800 sequences of fixed lengths between 15 and 250 words. In Fig. \ref{fig:3}, PRR peaks for sequences of 75–100 words. Yet, the performance degrades for shorter sequences due to limited semantic context exposed to the LM, and for longer sequences as the cumulative probability of a single unrecoverable error naturally increases.

We then shift our attention from sequence-level recovery accuracy to mask-level accuracy. We define Mask Recovery Accuracy (MRA) as the proportion of successfully resolved $\texttt{<mask>}$ tokens relative to the total number of masks generated. We evaluate MRA at sequence lengths of $N = 15$ and $N = 150$ to contrast scenarios with limited versus rich semantic context. For benchmark, Fig. \ref{fig:4} considers three possible methods: (1) LLR only, (2) T5Gemma only, and (3) the integrated LLR + T5Gemma approach (i.e., CL-SEC). While the LLR-only baseline struggles with channel noise and the T5Gemma-only model is hindered by semantic ambiguity,  CL-SEC significantly outperforms both, reaching a peak recovery rate of 44.6\% under the richer context of $N = 150$.

\section{Discussion and Conclusion}
This paper confirms the real-world viability of our previously proposed CL-SEC framework. Key takeaways of this paper include: 1) the successful real-time demonstration of the CL-SEC framework, which seamlessly fuses physical-layer soft information from conventional channel decoding with application-level semantics obtained via LMs to enhance decoding performance; 2) the design of a low-latency SDR middleware that efficiently bridges semantic processing at the application layer with raw hardware data at the physical layer, enabling stable real-time system operation; and 3) the design of a generalized LM interface that ensures CL-SEC's robust compatibility with modern encoder-decoder models. %Future work will focus on improving robustness against bursty noise and optimizing inference speeds for higher data rates.

\bibliographystyle{ACM-Reference-Format}
\bibliography{sample-base}

\end{document}